\documentclass[aps,pra,twocolumn,amsmath,amssymb,nofootinbib,showpacs,superscriptaddress]{revtex4-1}
\usepackage[english]{babel}
\usepackage{latexsym}
\usepackage{graphics}
\usepackage{graphicx}
\usepackage{epsfig}
\usepackage{color}
\usepackage{bm}
\usepackage{amsmath}
\usepackage{amssymb}
\usepackage{amsthm}
\usepackage{dcolumn}
\usepackage{bm}
\usepackage{float}
\usepackage{hyperref}
\usepackage{color}
\usepackage{epstopdf}
\usepackage{cleveref}
\usepackage[svgnames]{xcolor}
\usepackage{enumerate}
\usepackage{braket}
\hypersetup{hidelinks,colorlinks=true,allcolors=DarkBlue}

\theoremstyle{remark}
\newtheorem{remark}{Remark}

\begin{document}

\preprint{APS/123-QED}

\title{Practical issues in decoy-state quantum key distribution \\ based on the central limit theorem}

\author{A.S. Trushechkin}
\affiliation{Steklov Mathematical Institute of Russian Academy of Sciences, Moscow 119991, Russia}
\affiliation{National Research Nuclear University MEPhI, Moscow 115409, Russia}
\affiliation{Department of Mathematics and Russian Quantum Center, National University of Science and Technology MISiS, Moscow 119049, Russia}
\affiliation{Russian Quantum Center, Skolkovo, Moscow 143025, Russia}

\author{E.O. Kiktenko} 
\affiliation{Steklov Mathematical Institute of Russian Academy of Sciences, Moscow 119991, Russia}
\affiliation{Bauman Moscow State Technical University, Moscow 105005, Russia}
\affiliation{Russian Quantum Center, Skolkovo, Moscow 143025, Russia}
\affiliation{QApp, Skolkovo, Moscow 143025, Russia}

\author{A.K. Fedorov}
\affiliation{Department of Mathematics and Russian Quantum Center, National University of Science and Technology MISiS, Moscow 119049, Russia}
\affiliation{Russian Quantum Center, Skolkovo, Moscow 143025, Russia}
\affiliation{QApp, Skolkovo, Moscow 143025, Russia}

\date{\today}
\begin{abstract}
Decoy-state quantum key distribution is a standard tool for long-distance quantum communications. 
An important issue in this field is processing the decoy-state statistics taking into account statistical fluctuations (or ``finite-key effects''). 
In this work, we propose and analyze an option for decoy statistics processing, which is based on the central limit theorem.
We discuss such practical issues as inclusion of the failure probability of the decoy-states statistical estimates in the total failure probability of a QKD protocol 
and also taking into account the deviations of the binomially distributed random variables used in the estimations from the Gaussian distribution.
The results of numerical simulations show that the obtained estimations are quite tight.
The proposed technique can be used as a part of post-processing procedures for industrial quantum key distribution systems.
\end{abstract}
\maketitle

\section{Introduction}

Quantum key distribution (QKD) as the main part of quantum cryptography is known to provide information-theoretic (or unconditional) security of key distribution. 
However, QKD protocols like BB84 assume the employ of single-photon sources \cite{BB84}.~In contrast, 
real-life implementations of QKD setups are based on attenuated laser pulses instead of true single photons~\cite{Gisin,Scarani,LoRev,LoRev2}.~This 
realization makes QKD vulnerable to various attacks, such as the photon number splitting attack~\cite{Scarani,LoRev,LoRev2,Huttner,Brassard}.~A well-known tool 
for solving this problem is the decoy-state method, which can be considered as a standard technique used in many QKD realizations~\cite{Decoy01,Decoy02,Decoy03,Decoy04,Curty1,Curty2,Ma}. 
The decoy-state method uses laser pulses with different intensities. 
The intensities are chosen from a certain finite set. 
The choices for the pulses are kept in secret by the legitimate sender (Alice), 
but are publicly announced after the reception of all pulses by the legitimate receiver (Bob).
By analyzing (i) statistics of reception for pulses with different intensities and (ii) error rates for different intensities, 
one can estimate the fraction of single-photon pulses and the error rate for single-photon pulses. 
In particular, this allows detection of the photon number splitting attack~\cite{Decoy01,Decoy02,Decoy03,Decoy04,Curty1,Curty2,Ma}. 

An important task in the framework of the decoy-state QKD is to take into account statistical fluctuations (so called ``finite-key effects''). 
Several methods are proposed in the literature, including those based on the central limit theorem~\cite{Decoy04}, 
Chernoff--Hoefding method~\cite{Curty1,Curty2}, 
and improved Chernoff--Hoefding method \cite{Ma}. 

In the present work, we propose and analyze an option for processing decoy-state statistics based on the central limit theorem. 
Namely, we derive expressions for statistical estimations of the fraction of positions (bits) 
in the verified key obtained from single-photon pulses and the error rate in such positions. 
They are further used in calculations of the length of the secret (final) key with a given tolerable failure probability.
We also provide the results of numerical simulations, which show that these estimations are quite tight.
It should be mentioned that the methods based on the central limit theorem are criticized as not sufficiently rigorous~\cite{Ma}. 
However, we estimate the deviations from the Gaussian distribution in a rigorous way using the results of Ref.~\cite{ZubkovSerov}. 
Another important practical issue that we discuss is the accurate inclusion of the failure probability of the decoy states statistical estimates into the formula for the total failure probability.
We note that our analysis uses the decoy-state QKD protocol, which is described in Ref.~\cite{Decoy03}.

Our work is organized as follows.
In Sec.~\ref{sec:ppp}, we describe the basics of the QKD post-processing procedure.
In Sec.~\ref{sec:processing}, we present the suggested method for processing of decoy-state statistics based on the central limit theorem.
In Sec.~\ref{sec:simulation}, we use numerical simulations in order to compare the obtained estimations with theoretical limits of the decoy-state QKD protocol.
In Sec.~\ref{sec:deviations}, we estimate the deviations of the random variables used in our  processing procedure  from the Gaussian distribution.
We summarize the main results in Sec.~\ref{sec:conclusion}.

\section{QKD post-processing procedure}\label{sec:ppp}

The operating QKD protocol can be divided into several stages.
On the first quantum stage of a QKD protocol, Alice sends quantum states to Bob, who measures them. 
After the quantum stage Alice and Bob have two binary strings, the so-called \textit{raw keys}. 
The second stage is the use of post-processing procedures.
Let us recall the basic stages of post processing the raw keys for the BB84 QKD protocol 
(for details, see Refs.~\cite{Gisin,Scarani,LoRev,LoRev2,SECOQC,Gisin2,Fedorov}):

\begin{enumerate}[(i)]

\item \textit{Sifting}: Alice and Bob announce the bases they used for the preparation and measurement of quantum states and drop the positions with inconsistent bases from the raw keys. 
The resulting keys are called the \textit{sifted keys}. 
The decoy-states statistics is announced at this stage as well.

\item \textit{Information reconciliation}, also known as \textit{error correction}:
This entails removing discrepancies between Alice's and Bob's sifted keys via communication over the authenticated channel 
(for the last issues concerning the adaptation of error-correcting codes for QKD, see Ref.~\cite{Kiktenko2}).
Often this stage is completed by a verification procedure: one legitimate side send a hash-tag of his or her key to the other side to ensure the coincidence of their keys after the error correction. 
The blocks of the sifted keys which fail the verification test are discarded at this state.
The resulting common key is called the \textit{verified key}.
 
\item \textit{Parameter estimation}: 
this is the estimation of the quantum bit error rate (QBER) in the sifted keys. Also, processing the decoy states statistics is performed on this stage.

\item \textit{Privacy amplification}: 
the possible information obtained by an eavesdropper (Eve) about the keys is reduced to a negligible value
This is achieved by a special contraction of the verified key into a shorter key. 
For such contractions, 2-universal hash functions are used. 
This provides unconditional security against both classical and quantum eavesdroppers~\cite{ComposPA}. 
The resulting key is called the \textit{secret key} or the \textit{final key}. 
It is the output of a QKD protocol.

\end{enumerate}

Post-processing procedures require communication between Alice and Bob over a classical channel. 
This channel is not necessarily private (Eve is freely allowed to eavesdrop), but it must be authentic, 
i.e. Eve can neither change the messages sent via this channel nor send her own messages without being detected. 
To provide the authenticity of the classical channel, Alice and Bob uses message authentication codes. 
There are unconditionally secure message authentication codes~\cite{WegCar,Krawczyk,Krawczyk2}.

\begin{remark}\label{rem:pe}
Classically, the parameter estimation stage precedes the information reconciliation~\cite{Gisin}.
The QBER value is estimated by random sampling from the sifted keys (of course, being publicly announced this sample is discarded from the sifted keys). 
In our scheme, following Ref.~\cite{Gisin2}, the QBER value is determined after the information reconciliation and verification stages. 
Clearly, the straightforward comparison of the keys before and after the error correction procedure provides the exact number of corrected errors and corresponding QBER value. 
This allows one to avoid discarding a part of the sifted keys. 
This scheme was also used in the recently suggested symmetric blind information reconciliation method~\cite{Kiktenko2}. 
\end{remark}

The results of the statistical analysis of decoy states are used in the privacy amplification stage to calculate the length of the final key (contraction rate) which provides the required degree of security. For the BB84 protocol, the formula is as follows~\cite{Tomamichel,TomRenner}:
\begin{equation}\label{eq:pa}
	l_{\rm sec}=\hat\kappa_1^{\rm l} l_{\rm ver}[1-h(\hat e_1^{\rm u})]-\text{leak}_{\text{ec}}+5\log_2\varepsilon_{\rm pa},
\end{equation}
where $l_{\rm ver}$ is the length of the verified key, 
$\text{leak}_{\text{ec}}$ the amount of information (number of bits) about the sifted keys leaked to Eve during the information reconciliation stage,
\begin{equation}
	h(x)=-x\log_2x-(1-x)\log_2(1-x)
\end{equation} 
is the binary entropy function, and $\varepsilon_{\rm pa}$ is a tolerable failure probability for the privacy amplification stage. 
It can be interpreted as a probability that the privacy amplification stage has not destroyed all of Eve's information on the verified key and, so Eve has partial non-negligible information on the final key.

Further, $\hat\kappa_1^{\rm l}$ is a lower bound on the fraction of bits in the verified key obtained from single-photon pulses, 
and $\hat e_1^{\rm u}$ is an upper bound on the fraction of errors in such positions in the sifted keys. 
It is assumed that the bits of the verified keys obtained from multiphoton pulses are known to the eavesdropper. 
The quantity $h(\hat e_1^{\rm u})$ determines the potential knowledge of the bits obtained from single-photon pulses by the eavesdropper. 
This reflects the essence of QKD: it is impossible to get knowledge of the bits of the sifted key obtained from single-photon pulses without introducing errors in them. 
The estimation $\hat\kappa_1^{\rm l}$ and $\hat e_1^{\rm u}$ is the purpose of the decoy statistics analysis, which is given in the next section and is the main subject of the present paper. 
The failure probability for these estimates (the probability that at least one of these estimates is not true) must not be greater than some value $\varepsilon_{\rm decoy}$.

If $l_{\rm sec}$ given by Eq.~(\ref{eq:pa}) is positive, then the secret key distribution is possible. 
If $\tau$ is the time needed to generate a verified key with the length $l_{\rm ver}$, 
then the secret key rate can be defined as follows:
\begin{equation}
	R_{\rm sec}=l_{\rm sec}/\tau.
\end{equation} 

\begin{remark}
Let us comment on expression (\ref{eq:pa}). 
Essentially, it is taken from Ref.~\cite{Tomamichel}, where a rigorous proof of unconditional security of the BB84 protocol is given. 
Strictly speaking, the proof was given for the case when the QBER is estimated by random sampling from the sifted keys (see Remark~\ref{rem:pe}). 
However, the scheme when the QBER is estimated after the information reconciliation, simplifies the proof and formulas, since, 
in this case, the QBER is estimated not probabilistically but deterministically. 
Actually, the function $\varepsilon_{\rm pe}(\nu)$ in Theorem~3 in Ref.~\cite{Tomamichel}, 
which is the probability that the QBER estimation is incorrect, can be set to zero for all $\nu$.
\end{remark}

The total failure probability of the QKD system is the sum of the failure probabilities of each component: 
verification, authentication, privacy amplification, and statistical estimations of $\hat\kappa_1^{\rm l}$ and $\hat e_1^{\rm u}$:
\begin{equation}\label{eq:eps}
	\varepsilon_{\rm qkd}=\varepsilon_{\rm ver}+\varepsilon_{\rm aut}+\varepsilon_{\rm pa}+\varepsilon_{\rm decoy}.
\end{equation}
Here $\varepsilon_{\rm ver}$ is the probability that verification hash tags of Alice and Bob coincide, whereas their keys after the information reconciliation do not, 
and $\varepsilon_{\rm aut}$ is the probability that message authentication codes do not detect Eve's interference into the classical channel.

The meaning of the total failure probability $\varepsilon_{\rm qkd}$ is as follows: 
the key generated by the QKD protocol is indistinguishable from the perfectly secure key in any possible context (any possible application of this key) 
with the exception probability at most $\varepsilon_{\rm qkd}$ (see Ref.~\cite{Compos}).

\begin{remark}
More precisely, $\varepsilon_{\rm qkd}$ is the trace distance between the actual joint classical-quantum state of Alice, Bob, and Eve  and the ideal one, 
which corresponds to the case when Alice and Bob either have aborted the protocol or Alice's and Bob's keys coincide and are completely uncorrelated with Eve's state. 
But $\varepsilon_{\rm qkd}$ as a failure probability, the real state coincides with the ideal with the probability of at least $1-\varepsilon_{\rm qkd}$ \cite{ComposPA}. 
The derivation of formula (\ref{eq:eps}) is given in the Appendix.
\end{remark}

\begin{figure*}[t]
	\begin{center}
		\includegraphics[width=0.975\linewidth]{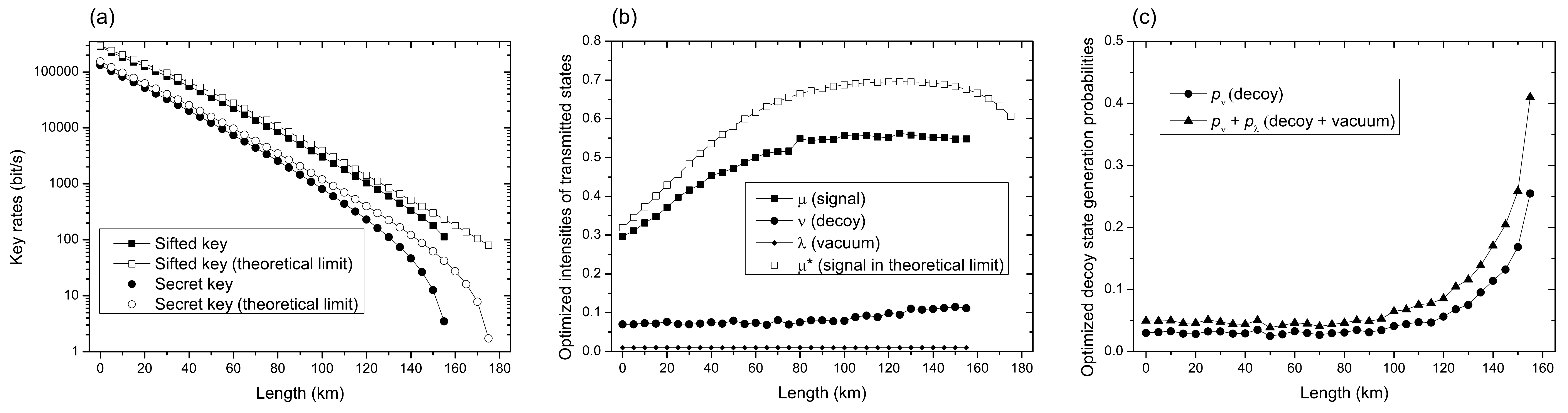}
	\end{center}
	\vskip -6mm
	\caption{Comparison of the proposed parameter estimation procedure with the theoretical limit for relevant parameters of the QKD setup.
		In (a) sifted and secret keys rates are presented as functions of the communication distance. 
		In (b) the optimized intensities for signal and decoy states, together with optimized intensity for signal states in the limit case, are shown as functions of the communication distance.
		In (c) the optimized fraction of decoy-states probabilities as functions of the communication distance are given. 
		We note that the fraction of signal states is given by $p_\mu=1-(p_{\nu}+p_\lambda)$.}
	\label{fig:simulation}
\end{figure*}

\section{Decoy-state statistics processing}\label{sec:processing}

Let us describe the estimations of $\hat\kappa_1^{\rm l}$ and $\hat e_1^{\rm u}$ used in Eq.~(\ref{eq:pa}). 
Here we adopt a finite-key version of the decoy statistics analysis described in Ref.~\cite{Decoy04}. 
Namely, in each round Alice sends to Bob a fixed number $N$ pulses. 
Each pulse has the ``signal intensity'' $\mu>0$ with the probability $p_\mu$, 
or the ``decoy intensity'' $\nu>0$ with the probability $p_\nu$, 
or the ``vacuum intensity'' $\lambda\geq0$ with the probability $p_\lambda=1-p_\mu-p_\nu$. 
We note that the intensity of the ``vacuum state'' $\lambda$ is close to zero, but not exactly zero due to the technical reasons.
In our consideration we assume  $\lambda=0.01$.
In fact, the ``vacuum intensity'' is the second decoy intensity. 
It is required that $\lambda<\nu/2$ and $\lambda+\nu<\mu$. 
Signal pulses are used to establish the raw key (and then the sifted, verified, and secret keys), whereas decoy pulses are used to estimate $\hat\kappa_1^{\rm l}$ and $\hat e_1^{\rm u}$.

Let $N$ be the total number of pulses sent by Alice, $N_\mu,N_\nu$, and $N_\lambda$ be the numbers of signal, decoy, and vacuum pulses sent by Alice 
(generally, they do not coincide with $p_\mu N$, $p_\nu N$, and $p_\lambda N$ due to statistical fluctuations, but, of course, $N_\mu+N_\nu+N_\lambda=N$),
and $n_\mu,n_\nu,n_\lambda$ be the numbers of the corresponding pulses registered by Bob.
Further, let $Q_\mu$ the probability that a signal pulse is registered by Bob,
$Q_\nu$ and $Q_\lambda$ the corresponding probabilities for decoy and vacuum pulses, 
and $Q_1$ be  the joint probability that a pulse contains a single photon and that it is registered. 
Then
\begin{equation}
	\theta_1=Q_1/Q_\mu
\end{equation}
is the probability that a bit in the sifted (as well as verified) key is obtained from a single-photon pulse. 
Finally, let $\kappa_1$ be the actual fraction of bits in the verified key obtained from single-photon pulses (it may differ from $\theta_1$ due to statistical fluctuations). 

We note that the random variables $N_\alpha$ ($\alpha\in\{\mu,\nu,\lambda\}$) and $l_{\rm ver}\kappa_1$ are binomially distributed. 
Indeed, each of $N$ pulses is, for example, a signal pulse with the probability $p_\mu$ independently of other pulses, and the number of the signal pulses $N_\mu$ is not fixed (random). 
Another probability distribution widely used in QKD is the hypergeometric distribution arising from sampling without replacement. 
Here (like, e.g., in Ref.~\cite{Ma}) we do not use sampling without replacement, 
but use the independent random choice scheme giving rise to the binomial distribution for the number of choices of a certain alternative (type of pulse).
 
If a random variable $X$ follows the binomial distribution with the number of experiments $n$ and the success probability in one experiment $p$, then we will write $X\sim{\rm Bi}(n,p)$. 
Then $N_\alpha\sim{\rm Bi}(N,p_\alpha)$ and $l_{\rm ver}\kappa_1\sim{\rm Bi}(l_{\rm ver},\theta_1)$. If the value of $N_\alpha$ is known and fixed (i.e., if we treat it as non-random), 
then $n_\alpha$ are also binomially distributed: $n_\alpha\sim{\rm Bi}(N_\alpha,Q_\alpha)$. 
Indeed, each pulse of a given type is detected with the probability $Q_\alpha$ independently of other pulses of this type.

In order to estimate $\kappa_1$ from below, we should estimate $\theta_1$ from below. To do this, we should estimate $Q_1$ from below and $Q_\mu$ from above.

According to Ref.~\cite{Decoy04}, the lower bound for $Q_1$ is as follows:
\begin{equation}\label{eq:q1}
\begin{split}
	Q_1&\geq\frac{\mu e^{-\mu}}{\nu\left(1-\frac\nu\mu\right)-\lambda\left(1-\frac\lambda\mu\right)}\times\\
	&\times\left[Q_\nu e^\nu-Q_\lambda e^\lambda-\frac{\nu^2-\lambda^2}{\mu^2}(Q_\mu e^\mu-Y_0^{\rm l})\right],
\end{split}
\end{equation}
where
\begin{equation}\label{eq:y0l}
	Y_0^{\rm l}=\max\left\lbrace\frac{\nu Q_\lambda e^\lambda-\lambda Q_\nu e^\nu}{\nu-\lambda},0\right\rbrace
\end{equation}
is the lower bound for the probability that Bob obtains a click event provided that the pulse contains no photons. 
The estimates of $Q_\mu,Q_\nu$ and $Q_\lambda$ are given by:
\begin{equation}
	\hat Q_\alpha=n_\alpha/N_\alpha, \quad \alpha=\mu,\nu,\lambda.
\end{equation}
Here and in the following the notation without a ``hat'' denotes a true value of a probability (a parameter in the binomial distribution), 
while the notation with a ``hat'' denotes its statistical estimate (i.e., a random variable).

Due to the central limit theorem, 
the distribution of the random variable $\hat Q_\alpha$ is well approximated by the normal distribution with the mean $Q_\alpha$ and standard deviation $\sqrt{{Q_\alpha(1-Q_\alpha)}/{N_\alpha}}$. 
If we denote 
\begin{equation}\label{eq:phi}
	\varphi=\Phi^{-1}\left(1-\frac{\varepsilon_{\rm decoy}}a\right),
\end{equation}
where $\Phi^{-1}$ is the quantile function for the standard normal distribution and $a$ is some constant to be specified, then
\begin{equation}
	P\left[Q_\alpha-\hat Q_\alpha\geq\varphi\sqrt{\frac{Q_\alpha(1-Q_\alpha)}{N_\alpha}}\right]\leq
\frac{\varepsilon_{\rm decoy}}a.
\end{equation}
This gives lower and upper bounds on $Q_\alpha$:
\begin{equation}\label{eq:Qj}
	\hat Q_\alpha^{\rm u,l}=\hat Q_\alpha\pm\varphi\sqrt{\frac{\hat Q_\alpha(1-\hat Q_\alpha)}{N_\alpha}}.
\end{equation}
Each bound is satisfied with the probability not less than $1-\varepsilon_{\rm decoy}/a$. 
We will need the upper bound on $Q_\mu$ and two-sided bounds on $Q_\nu$ and $Q_\lambda$. 
These five bounds are simultaneously satisfied with the probability not less than $1-5\varepsilon_{\rm decoy}/a$. 
Substitution of these bounds into Eq.~(\ref{eq:q1}) yields:
\begin{equation}
\begin{split}
	\!\!&Q_1\geq\frac{\mu e^{-\mu}}{\nu\left(1-\frac\nu\mu\right)-\lambda\left(1-\frac\lambda\mu\right)}\times\\
	\!\!&\times\left[\hat Q^{\rm l}_\nu e^\nu-\hat Q^{\rm u}_\lambda e^\lambda-\frac{\nu^2-\lambda^2}{\mu^2}(\hat Q^{\rm u}_\mu e^\mu-\hat Y_0^{\rm l})\right]\equiv \hat Q_1^{\rm l},
\end{split}
\end{equation}
where
\begin{equation}
\hat Y_0^{\rm l}=\max\left\lbrace
	\frac{\nu \hat Q^{\rm l}_\lambda e^\lambda-\lambda \hat Q^{\rm u}_\nu e^\nu}{\nu-\lambda},0\right\rbrace.
\end{equation}
Thus, we arrive at the following expression: 
\begin{equation}
	\theta_1\geq\frac{\hat Q^{\rm l}_1}{\hat Q^{\rm u}_\mu}=\hat\theta_1^{\rm l}
\end{equation}
with the probability not less than $1-5\varepsilon_{\rm decoy}/a$. 
The actual fraction $\kappa_1$ is estimated from below as 
\begin{equation}
	\kappa_1\geq\theta_1-\varphi\sqrt{\frac{\theta_1(1-\theta_1)}{l_{\rm ver}}}.
\end{equation}
with the probability not less than $1-\varepsilon_{\rm decoy}/a$, or,
\begin{equation}\label{eq:k1}
	\kappa_1\geq\hat\theta^{\rm l}_1-
	\varphi\sqrt{\frac{\hat\theta^{\rm l}_1(1-\hat\theta^{\rm l}_1)}{l_{\rm ver}}}
	\equiv\hat\kappa_1^{\rm l}.
\end{equation}
with the probability not less than $1-6\varepsilon_{\rm decoy}/a$. 
We have obtained one of two estimates participating in Eq.~(\ref{eq:pa}). 

Let us now find an upper bound for the error rate of the single-photon states. 
Though the formulas of these bounds are known (see Ref.~\cite{Decoy04}), the use of the binomial distribution should be analyzed in more detail. 
If Eve performs a coherent attack, then the errors in different positions of the keys cannot be treated as independent events. 
However, we are going to show that we can still use the binomial distribution. 

Let $n_i$ and $e_i$ denote the number of bits in the verified key obtained from the $i$-photon pulses and  the error rate in the $i$-photon states, respectively 
(i.e., $n_ie_i$ is the number of errors in the bits obtained from the $i$-photon pulses). 
Also let $e_\mu$ denotes the total error rate (QBER) for signal pulses. 
Then
\begin{eqnarray}
	l_{\rm ver} e_\mu&=&\sum_{i=0}^\infty n_ie_i\leq e_0n_0+e_1n_1, \\
	e_1 &\leq& \frac{l_{\rm ver} e_\mu-e_0n_0}{n_1}=
	\frac{e_\mu-e_0n_0/l_{\rm ver}}{\kappa_1},
\end{eqnarray}
where we have used $n_1=\kappa_1l_{\rm ver}$, by definition of $\kappa_1$. 
Obviously, the probability of error for a vacuum pulse is 1/2 for both natural noise and Eve's attack. 
Indeed, if there is no eavesdropping, then only dark counts can cause the click event on the Bob's side. 
If the Bob's detectors are identical, then they have equal probabilities of a click. 
Moreover, if they are memoryless (after a certain dead time), then the error events in vacuum pulses are independent from each other. 
Now consider the case of the presence of Eve. She has no way of knowing about the bit sent by Alice since the pulse contains no photons. 
The only thing she can do is to send her own pulse. 
But since she does not know the Alice's bit, her bit can be either correct or not with equal probabilities and independently of the correctness of other bits. 
Hence, $e_0n_0\sim{\rm Bi}(n_0,1/2)$ for a fixed $n_0$. But $n_0$ is also a binomially distributed random variable. 
For each of $N_\mu$ signal pulses, the joint probability that a signal pulse contains zero photons, 
the basis choices of Alice and Bob coincide, and Bob has a click event is $e^{-\mu}Y_0/2$ [see Eq.~(\ref{eq:y0l})]. 
Hence, $n_0\sim{\rm Bi}(N_\mu,e^{-\mu}Y_0/2)$, $e_0n_0\sim{\rm Bi}(N_\mu,e^{-\mu}Y_0/4)$,
\begin{equation}\label{eq:e0n0}
\begin{split}
	e_0n_0&\geq \frac{N_\mu e^{-\mu}Y_0^{\rm l}}4-
	\varphi\sqrt{N_\mu \frac{e^{-\mu}Y_0^{\rm l}}4
	\left(1-\frac{e^{-\mu}Y_0^{\rm l}}4\right)}\\
	&\equiv \upsilon,
\end{split}
\end{equation}
and
\begin{equation}\label{eq:e1}
e_1 \leq
\frac{e_\mu-\upsilon/l_{\rm ver}}{\hat\kappa_1^{\rm l}}
\equiv \hat e_1^{\rm u}
\end{equation}
with the probability not less than $1-\varepsilon_{\rm decoy}/a$. 

Thus, all statistical estimates are satisfied with the probability not less than $1-7\varepsilon_{\rm decoy}/a$. Since they must be satisfied with the probability not less than $1-\varepsilon_{\rm decoy}$, we set $a=7$.

\section{Simulation results}\label{sec:simulation}

We then consider realization of the described procedure on the realistic ``plug-and-play'' QKD setup~\cite{Sokolov} 
and compare the obtained results with theoretical limitations.
The parameters of the QKD setup implementation are as follows: 
number of pulses in train $5\times10^4$, 
repetition rate of pulses in train 300 MHz, 
storage line length 17~km,
detectors efficiency 10\%, 
detectors dead time 1~$\mu s$, 
dark count probability $3\times10^{-7}$, 
additional losses on Bob's side 5~dB, 
fiber attenuation coefficient 0.2~dB/km, 
and interference visibility 97\%.

\begin{figure}[t]
	\begin{center}
		\includegraphics[width=0.9\columnwidth]{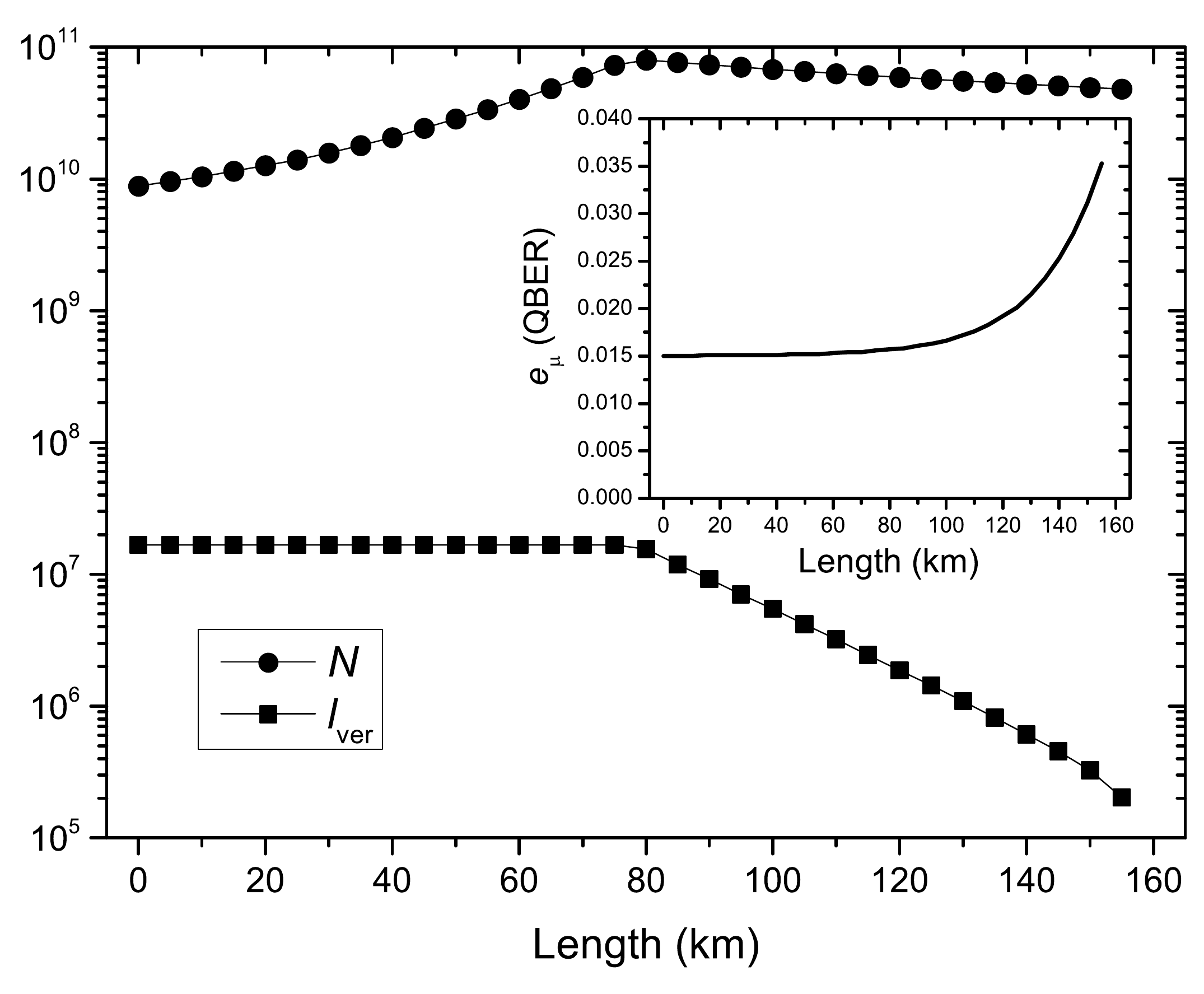}
	\end{center}
	\vskip -7mm
	\caption{Total number of transmitted pulses $N$ and length of verified key $l_{\rm ver}$ as functions of the communication distance. 
	In the inset the value of QBER is shown as a function of the communication distance.}
	\label{fig:qber}
\end{figure}

The parameters of the post-processing procedure are as follows: 
\begin{equation}
	\varepsilon_{\rm ver}=\varepsilon_{\rm aut}=\varepsilon_{\rm pa}=\varepsilon_{\rm decoy}=10^{-12}.
\end{equation}
The information leakage on the procedure of error correction and verification are~\cite{Kiktenko2}: 
\begin{equation}
	{\rm leak_{ec}}=f_{\rm ec}h(e_\mu), \quad f_{\rm ec}=1.15.
\end{equation}
The length of processed block $l_{\rm ver}$ is limited by the value of 16~Mbits or the length of sifted key accumulated after 30 min of the operation of the QKD setup.
We use the differential evolution method for numerical optimization of signal and decoy intensities ($\mu$ and $\nu$) together with their generation probabilities ($p_\mu$ and $p_\nu$).
The ``vacuum intensity'' is fixed at the level $\lambda=0.01$, and its generation probability is given by $p_\lambda=1-p_\mu-p_\nu$.
The results are presented in Fig.~\ref{fig:simulation} and Fig.~\ref{fig:qber}.

We then compare the results given by our approach with the theoretic limit, 
where we neglect statistical fluctuations and assume that we know the exact values of $\kappa_1$ and $e_1$
(i.e., there is no need in statistical estimates and decoy states, $p_\mu=1$, $p_\nu=p_\lambda=0$).
In the theoretical limit, the secret key rate is given by the expression: 
\begin{equation}\label{eq:R_th}
	R_{\rm sec}^{*}=R_{\rm sift}\{\kappa_1[1-h(e_1)]-f_{\rm ec}h(e_\mu)\},
\end{equation}
where $R_{\rm sift}$ is the sifted key rate. 
The quantities $R_{\rm sift}$, $\kappa_1$, $e_1$, and $e_\mu$ in Eq.~(\ref{eq:R_th}) depend on the intensity of signal pulses ($\mu^*$ for the theoretical limit case).
These quantities are obtained from the numerical optimization of the intensities for various communication distances.

The results of the comparison of our approach and theoretical limit are presented in Fig.~\ref{fig:simulation}.
In Fig.~\ref{fig:simulation}(a) it is shown that the proposed approach rather closely approximates the theoretical limit on distances up to 100--120 km.
The optimal operating of post-processing procedures on such distances is important, in particular, for inter-city QKD for future quantum networks.

We note that the sifted key rate in the theoretic limit is higher due to higher optimal signal intensity [Fig. \ref{fig:simulation}(b)] and absence of decoy states.
Also note that the optimal fraction of decoy states in the proposed approach is relatively small and is about $5\%$ for distances less than 100 km [see Fig. \ref{fig:simulation}(c)].

\section{Deviations from the Gaussian distribution}\label{sec:deviations}

In the discussed approach the statistical fluctuations are treated in the framework of the central limit theorem, i.e. we assume that the binomially distributed random variables 
(the fraction of positions obtained by single-photon pulses $\kappa_1$ and the error rate in these positions $e_1$) are well approximated by the Gaussian distribution. 
This approach is criticized as not sufficiently rigorous \cite{Ma}, in contrast to other approaches which are more rigorous but give slightly worse estimates.  
Now we are going to estimate deviations from the Gaussian distribution.

Let $X\sim{\rm Bi}(n,p)$. 
Then according to Ref.~\cite{ZubkovSerov}:
\begin{equation}\label{eq:zubkovserov}
	C_{n,p}(k)\leq\Pr[X\leq k]\leq C_{n,p}(k+1),
\end{equation}
where $C_{n,p}(0)=(1-p)^n$, $C_{n,p}(n)=1-p^n$, and
\begin{equation}
	C_{n,p}(k)=\Phi\left({\rm sgn}\left(\tfrac kn-p\right)\sqrt{2nH\left(\tfrac kn,p\right)}\right)
\end{equation}
for $k=1,\ldots,n-1$,
\begin{equation}
	\Phi(x)=\frac1{\sqrt{2\pi}}\int_{-\infty}^x e^{-t^2/2}dt,
\end{equation}
\begin{equation}
	H(x,p)=x\ln\frac xp+(1-x)\ln\frac{1-x}{1-p},
\end{equation}
and ${\rm sgn}(x)=x/{|x|}$ for $x\neq0$ and ${\rm sgn}(0)=0$. 

If $k=np+\varphi\sqrt{np(1-p)}$, then it is straightforward to show using Taylor's theorem (for $n\to\infty$) that
\begin{subequations}\label{eq:devia}
\begin{eqnarray}
	&&C_{n,p}(k)=\Phi(\varphi)-
	\frac{\varphi^2(1-2p)e^{-\varphi^2/2}}{6\sqrt{2\pi np(1-p)}}+O(\tfrac1n),\\
	&&C_{n,p}(k+1)=\Phi(\varphi)-
	\frac{\varphi^2(1-2p)e^{-\varphi^2/2}}{6\sqrt{2\pi np(1-p)}}\nonumber\\
	&&\qquad\qquad\quad\:+\frac{e^{-\varphi^2/2}}{\sqrt{2\pi np(1-p)}}+O(\tfrac1n).
\end{eqnarray}
\end{subequations}
We see that the deviations from the Gaussian distribution become significant for small $n$ or $p$ close to 0 or 1.

In Equations~(\ref{eq:Qj}), (\ref{eq:k1}), and (\ref{eq:e0n0}) we took $\varphi$ such that
\begin{equation}
\Phi(\varphi)=1-\frac{\varepsilon_{\rm decoy}}7
\end{equation}
[see Eq.~(\ref{eq:phi})] for $\varepsilon_{\rm decoy}=10^{-12}$, i.e., $\varphi\approx7.30$. But the precise value $\varepsilon'_{\rm decoy}$ of the failure probability is given by
\begin{equation}
\Pr[X\leq np+\varphi\sqrt{np(1-p)}]=1-\frac{\varepsilon'_{\rm decoy}}7.
\end{equation}
Eqs. (\ref{eq:devia}) can be used to estimate the difference between the precise value $\varepsilon'_{\rm decoy}$ and the required value $\varepsilon_{\rm decoy}$. 

Let us consider the worst-case scenarios for estimates~(\ref{eq:Qj}), (\ref{eq:k1}), and (\ref{eq:e0n0}). 
In Eq.~(\ref{eq:Qj}), the minimal possible $N_\alpha$ is $N_\lambda\sim 10^8$ (when the length is close to zero) and all $\hat Q_\alpha$ are at least of the order $10^{-7}$ 
(they are bounded from below by dark count probability with a dead time corrections). 
The substitution of these worst-case parameters $n=10^8$ and $p=10^{-7}$ to Eq.~(\ref{eq:devia}) gives
\begin{equation}
	\varepsilon'_{\rm decoy}-\varepsilon_{\rm decoy}\approx 2\cdot10^{-15}
	\ll\varepsilon_{\rm decoy}=10^{-12}.
\end{equation} 
In Eq.~(\ref{eq:k1}), the worst-case parameters are $n=l_{\rm ver}=10^5$ and $p=\theta_1\approx0.47$ (corresponding to the maximal length 155~km):
\begin{equation}
	\varepsilon'_{\rm decoy}-\varepsilon_{\rm decoy}\approx 1.5\cdot10^{-16}
	\ll\varepsilon_{\rm decoy}=10^{-12}.
\end{equation} 
In Eq.~(\ref{eq:e0n0}), we always have $Y_0^{\rm l}=0$, so, in fact, we do not perform the statistical estimation and use a trivial estimate $n_0e_0\geq0$.  
We see that the precise value $\varepsilon'_{\rm decoy}$ may exceed the required value $\varepsilon_{\rm decoy}$ only by a negligible quantity. 
The higher-order corrections to the Gaussian distribution with respect to $n^{-1/2}$ are even smaller. 
Thus, for practical parameters one can use the proposed formulas based on the Gaussian distribution.

Derivation of a general procedure for decoy state statistics processing based on rigorous formula (\ref{eq:zubkovserov}) will be a subject for a subsequent work.

\section{Concluding remarks}\label{sec:conclusion}

We have presented a sort of the decoy state statistics processing. 
The final formulas are (\ref{eq:k1}) and (\ref{eq:e1}), which give the statistical estimates used in the formula for the length of the final secret key (\ref{eq:pa}).

Also we claim that the failure probability $\varepsilon_{\rm decoy}$ for the decoy states statistical estimates should be treated as an additional term in the total failure probability 
$\varepsilon_{\rm qkd}$ in Eq.~(\ref{eq:eps}). 
Usually, one simply puts $\varepsilon_{\rm decoy}=\varepsilon_{\rm pa}$ and not treat it as an additional term in the total failure probability. 
From the point of view of the rigorous theory, this is not correct: formula (\ref{eq:pa}) provides the failure probability at most $\varepsilon_{\rm pa}$ only for true single-photon sources or, 
at least, when we can estimate $\hat\kappa_1^{\rm l}$ and $\hat e_1^{\rm u}$ with certainty. 
If we cannot estimate these quantities with certainty, the failure probability of the estimate should be included as an additional term in the the total failure probability of the QKD system. 
The appendix is devoted to the rigorous derivation of Eq.~(\ref{eq:eps}) for the total failure probability.

Finally, we have shown that, for practical parameters, 
deviations of the binomially distributed random variables used in the decoy states statistics processing from the Gaussian distribution can be neglected.

The suggested option for the decoy-state processing is implemented in the proof-of-principle realization of the post-processing procedure for industrial QKD systems~\cite{Fedorov}, 
which is freely available under GNU general public license (GPL)~\cite{Code}.
This procedure is used in the modular QKD setup described in Ref.~\cite{Sokolov}. 

\section*{Acknowledgments}

We are grateful to Y.V.~Kurochkin, A.S.~Sokolov, and A.V.~Miller for useful discussions. 
The work of A.S.T. and E.O.K. was supported by the grant of the President of the Russian Federation (project MK-2815.2017.1).
A.K.F. is supported by the RFBR grant (17-08-00742).

\setcounter{equation}{0}
\setcounter{section}{0}
\renewcommand{\theequation}{A\arabic{equation}}

\section*{Appendix. Derivation of the formula for the failure probability}\label{sec:eps}

This section is devoted to the derivation of formula (\ref{eq:eps}). 
According to the results of Ref.~\cite{Portmann}:
\begin{equation}
	\varepsilon_{\rm qkd}=\varepsilon_{\rm corr}+\varepsilon_{\rm sec},
\end{equation}
where $\varepsilon_{\rm corr}$ and $\varepsilon_{\rm sec}$ stand for correctness (coincidence of Alice's and Bob's final keys) and secrecy (ignorance of Eve about the final key), respectively. 
Namely, $\varepsilon_{\rm corr}$ is the probability that Alice's and Bob's keys do not coincide, 
but the protocol was not aborted, and $\varepsilon_{\rm sec}$ is the trace distance between the actual joint classical-quantum state of Alice and Eve and the ideal one. In our case, 
\begin{equation}
	\varepsilon_{\rm corr}=\varepsilon_{\rm ver}+\varepsilon_{\rm aut},
\end{equation}
i.e., the Alice's and Bob's final keys coincide in the case of coincidence of their verification hash tags and their authentication hash tags
(otherwise Eve could interfere into their communication and fake the verification tags). 
So, the failure probability for correctness is the sum of failure probabilities of the verification and authentication.

Further, in the case of single-photon sources, $\varepsilon_{\rm sec}=\varepsilon_{\rm pa}$ (see, Ref.~\cite{Tomamichel}). 
In this case we can estimate $\hat\kappa_1^{\rm l}$ and $\hat e_1^{\rm u}$ in Eq.~(\ref{eq:pa}) with certainty.
If we cannot estimate these quantities with certainty, the failure probability of the estimate should be included as an additional term in the  total failure probability of the QKD system:
\begin{equation}\label{eq:epsec}
	\varepsilon_{\rm sec}=\varepsilon_{\rm pa}+\varepsilon_{\rm decoy}.
\end{equation}
A mathematical fact justifying expression Eq.~(\ref{eq:epsec}) is as follows. 
Consider two classical-quantum states:
\begin{align}
	\rho_{XY}&=\sum_{x\in\mathcal X} p_x\ket x\bra x\otimes\rho_{Y|x},\\
	\sigma_{XY}&=\sum_{x\in\mathcal X} p_x\ket x\bra x\otimes\sigma_{Y|x},
\end{align}
where $\mathcal X$ is a finite set, $p_x$ are probabilities. 
Let, further, $\Omega\subset X$ (event), $\overline\Omega=\mathcal X\backslash\Omega$, $p(\Omega)=\sum_{x\in\Omega}p_x$,  
\begin{align}
	\rho_{XE|\Omega}&=\frac1{p(\Omega)}\sum_{x\in\Omega} p_x\ket x\bra x\otimes\rho_{E|x},
	\\
	\sigma_{XE|\Omega}&=\frac1{p(\Omega)}\sum_{x\in\Omega} p_x\ket x\bra x\otimes\sigma_{E|x},
\end{align}
\begin{align}
	\rho_{XE}&=p(\Omega)\rho_{XE|\Omega}+(1-p(\Omega))\rho_{XE|\overline\Omega},
	\\
	\sigma_{XE}&=p(\Omega)\sigma_{XE|\Omega}+(1-p(\Omega))\sigma_{XE|\overline\Omega},
\end{align}
\begin{equation}\label{eq:dcond}
	D(\rho_{XE|\Omega},\sigma_{XE|\Omega})\leq\varepsilon_1,
\end{equation}
where $D$ is the trace distance (for details, see Refs.~\cite{Compos,ComposPA}) and $p(\Omega)\geq1-\varepsilon_2$. 
Then
\begin{equation}\label{eq:d}
\begin{split}
	D(\rho_{XE},\sigma_{XE})&=
	p(\Omega)D(\rho_{XE|\Omega},\sigma_{XE|\Omega})\\&+
	(1-p(\Omega))D(\rho_{XE|\overline\Omega},\sigma_{XE|\overline\Omega})\\
	&\leq(1-\varepsilon_2)\varepsilon_1+\varepsilon_2\\
	&\leq\varepsilon_1+\varepsilon_2,
\end{split}
\end{equation}
where we have used that the trace distance does not exceed unity. 
In our case $\mathcal X$ is the set of all pairs $(\kappa_1,e_1)$, $\Omega$ is the subset corresponding to the event 
\begin{equation}
	(\kappa_1\geq\hat\kappa_1^{\rm l}\text{ and }e_1\leq\hat e_1^{\rm u}),
\end{equation}
$\varepsilon_1=\varepsilon_{\rm pa}$, $\varepsilon_2=\varepsilon_{\rm decoy}$, 
(\ref{eq:dcond}) is the trace distance between the actual and the ideal final states of the protocol conditioned on the event that the statistical estimates of $\kappa_1$ and $e_1$ are true, 
and, finally, (\ref{eq:d}) is the total trace distance between the actual and the ideal final states of the protocol.

\begin{remark}
In Ref.~\cite{Fung}, another formula relating the failure probability and trace distance is used: 
if $\varepsilon$ is the failure probability, then the trace distance between the actual and the ideal state is bounded from above by $\sqrt{\varepsilon(2-\varepsilon)}$, 
instead of linear formulas (\ref{eq:eps}) and (\ref{eq:epsec}). 
The reason is the difference between the techniques of security proofs. 
In Ref.~\cite{Fung}, entanglement-distillation technique was assumed. 
Its final result is expressed in terms of fidelity $F(\rho,\sigma)$ between the actual joint state of Alice, Bob, 
and Eve $\rho$ and the ideal one $\sigma$. If $F(\rho,\sigma)\geq1-\varepsilon$, then the trace distance is bounded by
\begin{equation}
	D(\rho,\sigma)\leq\sqrt{1-F(\rho,\sigma)^2}\leq\sqrt{\varepsilon(2-\varepsilon)}.
\end{equation}
In contrast, proofs of Refs.~\cite{Tomamichel,TomRenner,Renner} are information-theoretic and their results are direct bounds on the trace distance 
(the leftover hash lemma is essentially the main ingredient yielding such a If we cannot estimatebound).
\end{remark}

\end{document}